\documentclass[slac_one]{revtex4}
\usepackage{graphicx}
\usepackage{fancyhdr}
\def \jinst#1#2#3{JINST\ {\bf#1}, #2, #3}

\begin{document}

\title{Alignment of the ATLAS Inner Detector Tracking System}

%

\author{M\"uge Karag\"oz \"Unel}
\thanks{On behalf of ATLAS Inner Detector Collaboration. 
The author thanks the ATLAS Inner Detector group members.}
\affiliation{University of Oxford, Denys Wilkinson Building, Keble Road, OX1 3RH, UK}

\begin{abstract}
The ATLAS detector at CERN's Large Hadron Collider (LHC) is equipped with a tracking
system at its core (the Inner Detector, ID) consisting of silicon and
gaseous straw tube detectors. 
The physics
performance of the ID requires a precision alignment; a
challenge involving complex algorithms and significant
computing power. 
The alignment algorithms were already validated on:
Combined Test Beam data, Cosmic Ray runs and simulated physics
events.
The alignment chain was tested on a daily basis
in exercises that mimicked ATLAS data taking
operations. ID commissioning after final installation into the ATLAS detector 
has yielded
thousands of reconstructed cosmic ray tracks, which have been 
used for an initial alignment of the ID before the LHC start-up.
A hardware system using Frequency Scanning
Interferometry will be used to monitor structural deformations. 
Given the programme outlined here, the ATLAS Inner Detector has had a solid preparation for LHC collisions.
\end{abstract}

\maketitle

\thispagestyle{fancy}

\section{INTRODUCTION} 

The 7.0m-long, 2.3m-high ATLAS Inner Detector (ID) consists of three main subsystems: Pixel, SemiConductor Tracker (SCT) and Transition Radiation Tracker (TRT), in radially increasing order. The ID is designed to measure momentum and charge of particle tracks and to determine vertices efficiently and precisely. 
It must be aligned with reliable accuracy and precision to meet the design specification~\cite{ID}.
The impact on physics performance of relative displacements between elements within the three main subsystems differs according to the direction of displacements, 
with the most sensitive being the $r\phi$ coordinate of the silicon (Si) detectors.

Achieving good alignment depends on: a good detector design and precision surveying during construction, 
effective software and hardware alignment algorithms and their efficient operation within 
daily data taking, understanding of systematic deformations, sufficient and prompt monitoring and validation. 
This note summarises the ATLAS ID alignment with selected illustrative examples of its performance.
	   
Table~\ref{id_summary} summarises the properties of the ID~\cite{det_paper}.
Initial positions of the detector elements were obtained during the construction and assembly surveys. The mounting and survey precision for the Si modules was less than 50$ \mu$m; for TRT wires, it was $O(100~\mu\rm{m})$.
For each tracker element; both Si and TRT modules, there are three
rotational and three translational degrees of freedom (DoF).  All six are
determined, except for TRT barrel modules, where translation along readout wires is neglected.

\begin{table}[t]
\begin{center}
\caption{Summary of components, DoF, intrinsic resolutions and estimated tolerances for alignment contribution to resolution~\cite{ID, det_paper} for the ATLAS ID.}
\vspace{2mm}
\begin{tabular}{|c|c|c|c|c|c|c|}
\hline 
\multicolumn{1}{|c}{\textbf{Subdetector}} & \multicolumn{2}{|c}{\textbf{Pixel}} & \multicolumn{2}{|c}{\textbf{SCT}} & \multicolumn{2}{|c|}{\textbf{TRT}} \\ \hline
\textbf{Components}  & Barrel & Endcap & Barrel & Endcap & ~~Barrel~~ & ~Endcap~ \\ \hline
\multicolumn{1}{|c}{\textbf{Unit Cell}} & \multicolumn{2}{|c}{silicon pixel} & \multicolumn{2}{|c}{silicon strip} & \multicolumn{2}{|c|}{gaseous drift tube} \\ \hline
\multicolumn{1}{|c}{\textbf{Cell Dim.}} & \multicolumn{2}{|c}{50$\times$400 $\mu$m$^2$} & \multicolumn{1}{|c}{pitch: 80 $\mu$m} &\multicolumn{1}{|c}{pitch: 57-90 $\mu$m} & \multicolumn{2}{|c|}{diameter: 4 mm} \\ \hline
\textbf{Layers/Disks}  &  3 & 2$\times$3 & 4 & 2$\times$9 & 3 & 2$\times$14\footnote{In the near future, the granularity of the TRT endcap disks will be increased to 2$\times$40.}
\\ \hline
\textbf{Modules}  &  1456 & 2$\times$144 & 2112 & 2$\times$988 & 96  & 2$\times$398\\ \hline
\multicolumn{1}{|c}{\textbf{Total}} & \multicolumn{2}{|c}{1744} & \multicolumn{2}{|c}{4088} & \multicolumn{2}{|c|}{992} \\ \hline
\multicolumn{1}{|c}{\textbf{Total DoF}} & \multicolumn{2}{|c}{10464} & \multicolumn{2}{|c}{24528} & \multicolumn{2}{|c|}{5952} \\ \hline
\multicolumn{1}{|c}{\textbf{Resolutions}} & \multicolumn{2}{|c}{14 $\mu$m ($r\phi$)} & \multicolumn{2}{|c}{23 $\mu$m ($r\phi$)} & \multicolumn{2}{|c|}{130 $\mu$m} \\ 
\multicolumn{1}{|c}{\textbf{}} & \multicolumn{2}{|c}{115 (60\footnote{This is an improvement over the intrinsic resolution using a charge-sharing technique.}) $\mu$m ($r/z$)} & \multicolumn{2}{|c}{580 $\mu$m ($r/z$)} & \multicolumn{2}{|c|}{($\perp$ readout wire)} \\ \hline
\multicolumn{1}{|c}{\textbf{Tolerances}} & \multicolumn{2}{|c}{7 $\mu$m ($r\phi$)} & \multicolumn{2}{|c}{12 $\mu$m ($r\phi$)} & \multicolumn{2}{|c|}{30 $\mu$m ($r\phi$)} \\ \hline
\multicolumn{1}{|c}{\textbf{ }}  & \multicolumn{1}{|c}{20 $\mu$m ($z$)} & \multicolumn{1}{|c}{100 $\mu$m ($z$)} & \multicolumn{1}{|c}{50 $\mu$m ($z$)} & \multicolumn{1}{|c}{200 $\mu$m ($z$)} &  \multicolumn{2}{|c|}{N/A} \\
\hline
\end{tabular}
\label{id_summary}
\end{center}
\end{table}

Due to the size and granularity of the subdetectors and the operational demands placed upon them, the ID alignment is a numerical and computational 
challenge taken up using state-of-the-art tools and algorithms.
ATLAS uses track-based and optical alignment approaches to align the ID during ATLAS operation.
Track-based alignment within the ATLAS software framework relies on optimising residuals for particle tracks, where the residual is defined as the (2 or 3D) distance between the fitted track and the track hit recorded by a module. 
Currently, four independent track-based algorithms (three for Si and one for TRT) are implemented~\cite{det_paper}, of which 
Global $\chi^2$ and TRT $\chi^2$ are the current official baseline algorithms. 
All except Robust Alignment, are based on $\chi^2$ minimisation. Algorithms run iteratively to optimise the track-hit residuals by re-reconstructing tracks after each iteration.
Global $\chi^2$ algorithm minimises the $\chi^2$ from a simultaneous fit to
all track and alignment parameters, taking into account correlation between modules. 
When solving for the entire Si detector, there are 34992 DoF to consider, which is computationally demanding. TRT alignment is performed with respect to and after Silicon alignment in the alignment chain.


\section{PERFORMANCE}

ID alignment algorithms were validated and used on various datasets over the past few years. The first data challenge consisted of aligning a single radial tower of the ID in a Combined Test Beam setup~\cite{CTB_align}. 

\indent {\em \bf Cosmic Ray Commissioning:}
All Inner Detector subcomponents collected cosmic ray data in so-called SR1 tests before being lowered into the cavern~\cite{sr1}. 
A subsection of TRT and SCT barrels took combined data in June 2006 which was successfully used for an initial TRT-SCT global alignment and detector internal alignment in the absence of a magnetic field. 
For the 468 modules read out from the SCT in about 200k events, a residual width of 32 $\mu$m, which was within 1 $\mu$m of the ideal alignment in simulation, was reached. 
This was obtained starting from an unaligned detector with a residual width of 65~$\mu$m, which indicated an already good SCT barrel assembly precision.

\begin{figure}
\includegraphics[width=64mm]{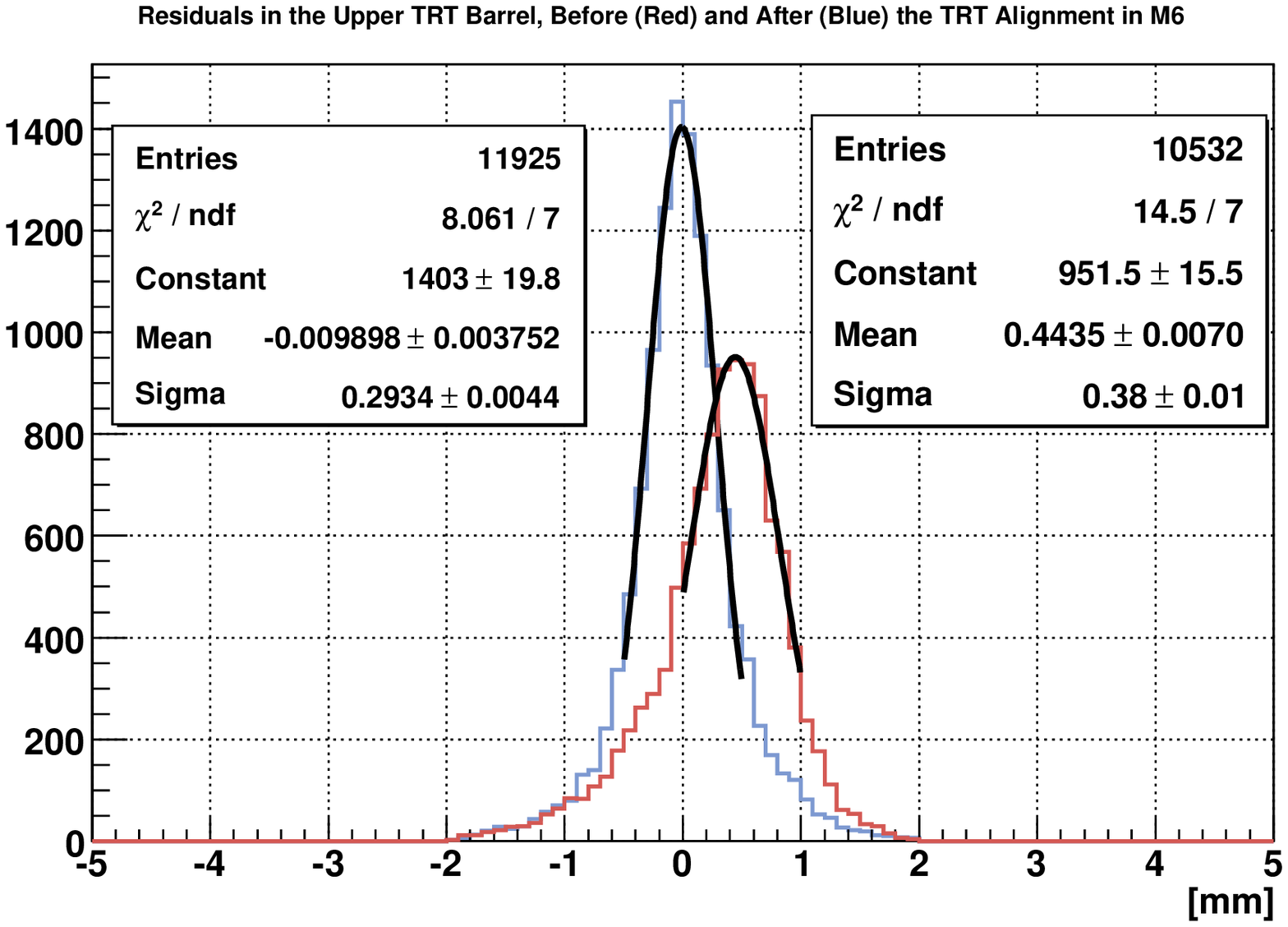}%
\includegraphics[width=46mm]{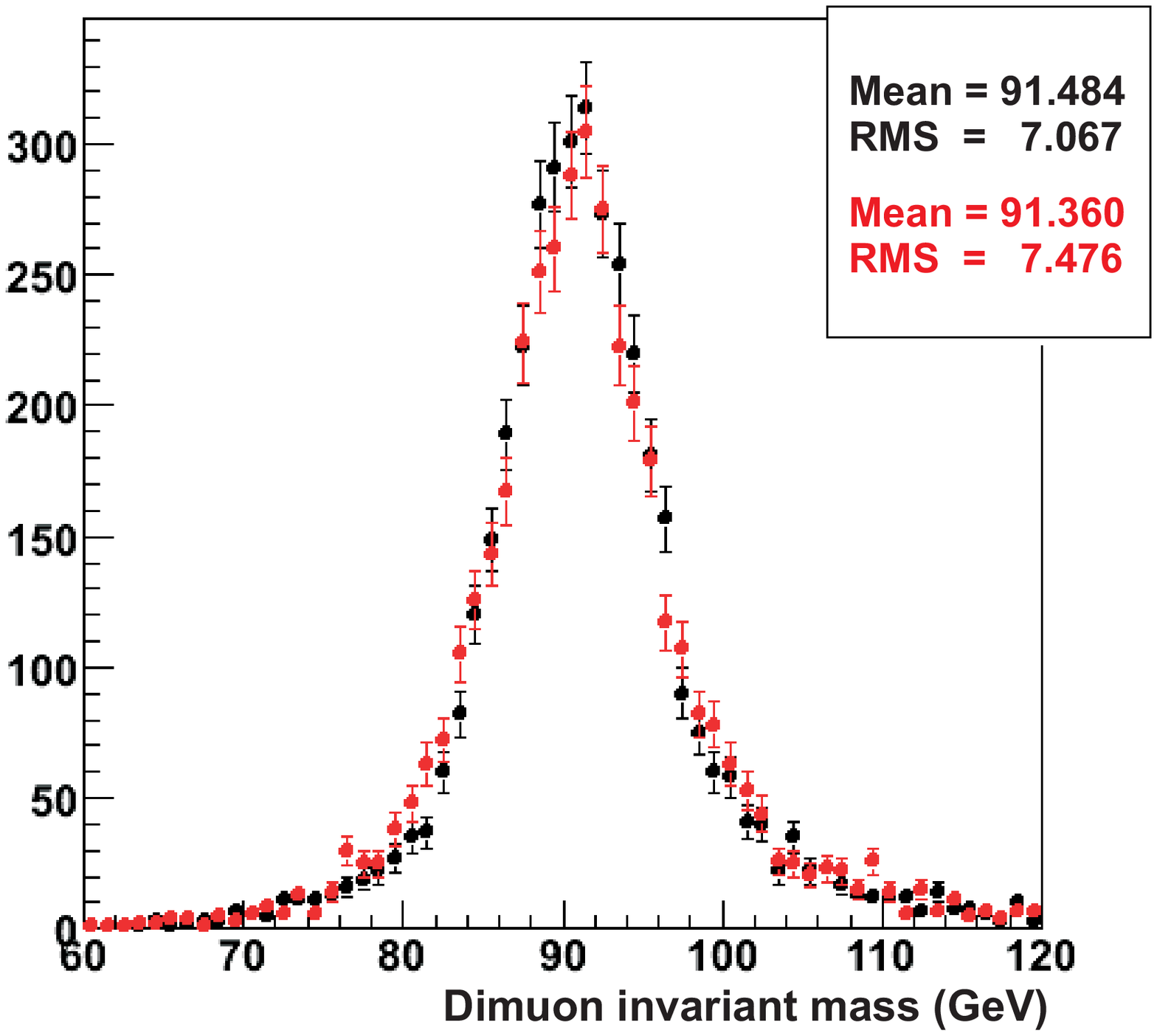}%
\vspace{-1mm}
\caption{\label{cosmic_results} Left: TRT residuals before (right peak) and after (left peak) after M6 alignment. Right: Invariant mass of muon pairs of $Z$ bosons in simulation showing the agreement between the FDR exercises (red points) and CSC alignment (black points) for Si-only tracks.}
\end{figure}

The first ID alignment in the ATLAS cavern (without the pixel detector) came from the cosmic runs (``M6 milestone" run) in March 2008, where about 12k usable events were collected. Five thousand tracks for SCT and 4k tracks for TRT were used to align the parts of the detectors covered by the trigger acceptance. 
Given the low number of events, a module level alignment was not attempted for the SCT. Nevertheless, alignment based on M6 data improved the knowledge of the module positions in both detectors with TRT residuals improving greatly (Fig.~\ref{cosmic_results}). 
The alignment information from M6 is now being used for the reconstruction of the cosmic data taken in the cavern.\footnote{At the time of writing, a combined Pixel, SCT and TRT alignment has been performed with recent cosmic data in the cavern.}

\indent {\em \bf Software and Operation Model Commissioning:}
Within the CSC (Computer System Commissioning) exercise, a realistic simulation of the ATLAS detector 
accounting for the assembly imperfections and material description was performed.
ID misalignments were introduced at 3 levels of detector structure hierarchy, 
according to the build precision: Level 1, entire subdetectors 
defined by mechanical freedom, $O(\rm{mm})$ translations and $O(\rm{mrad})$ rotations, Level 2, 
layers and disks for Si, barrel modules for TRT, $O(100~\mu\rm{m})$, and Level 3, modules, full DoF for Si, none for TRT, $O(100~\mu\rm{m})$. 
A ``multi-muon" sample, flatly distributed in track parameters, and a cosmic ray sample were used in sequence to produce one set of alignment constants which was validated using 
Z$\rightarrow \mu\mu$ samples.
The $Z$ mass peak was successfully recovered with current algorithms, however, residual biases in track parameters were observed, due to the so-called ``weak mode" deformations. In addition to constraints imposed by using cosmic data, other constraints are needed for these weak modes to reach ultimate precision. 
CSC result details are given elsewhere~\cite{det_paper}. 

\begin{figure}
\includegraphics[width=88mm]{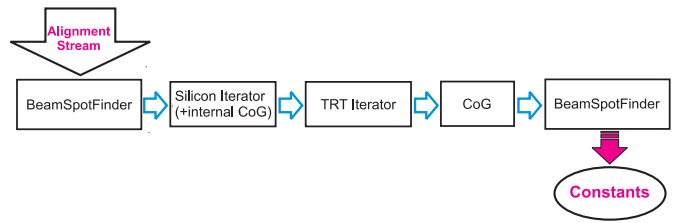}%
\vspace{-1mm}
\caption{\label{flow}FDR2 alignment operation loop to produce daily alignment constants. CoG stands for Centre of Gravity, which is a step to correct the global coordinates of the ID after alignment.}
\end{figure}

\begin{figure}
\includegraphics[width=65mm]{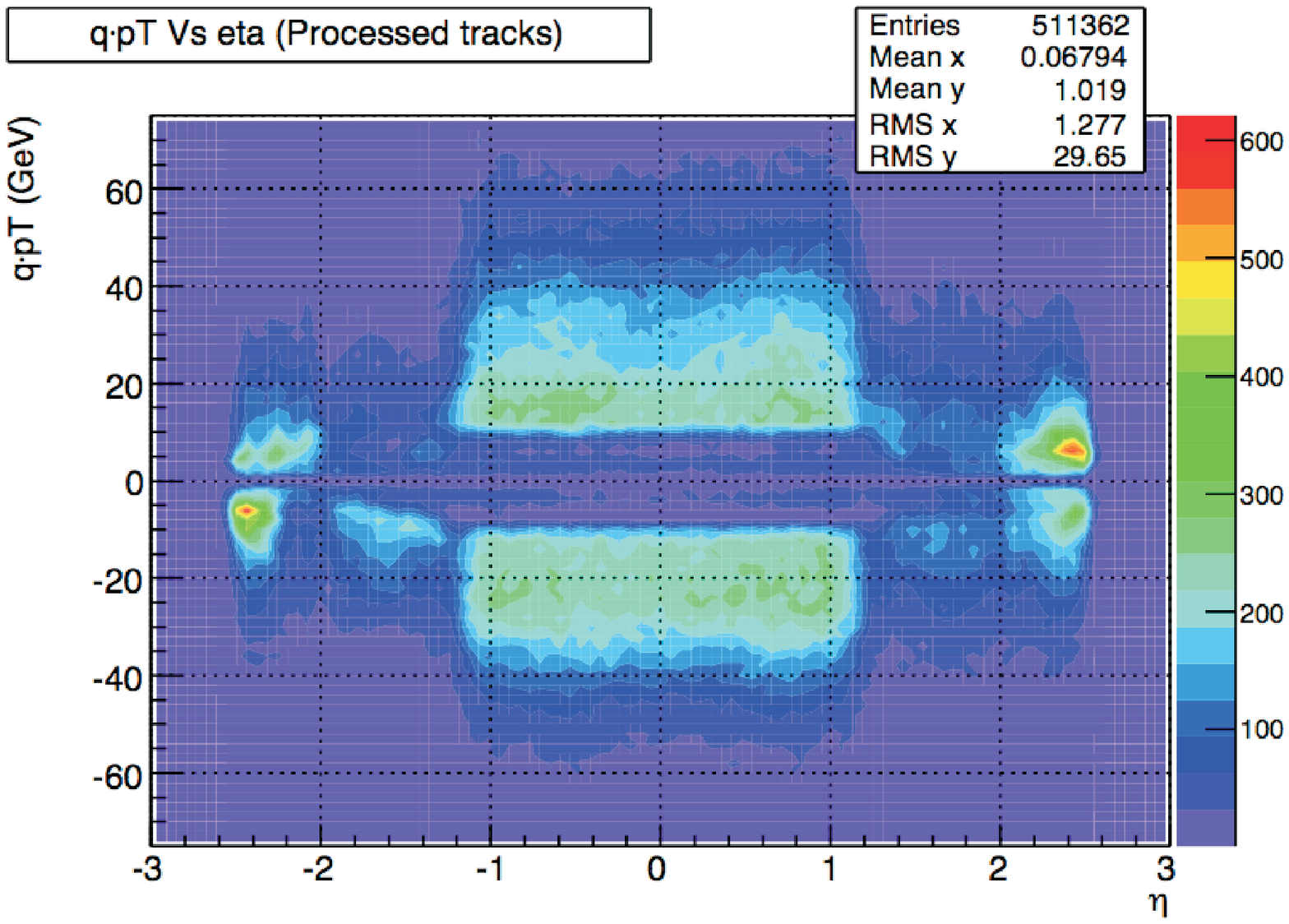}%
\includegraphics[width=64mm]{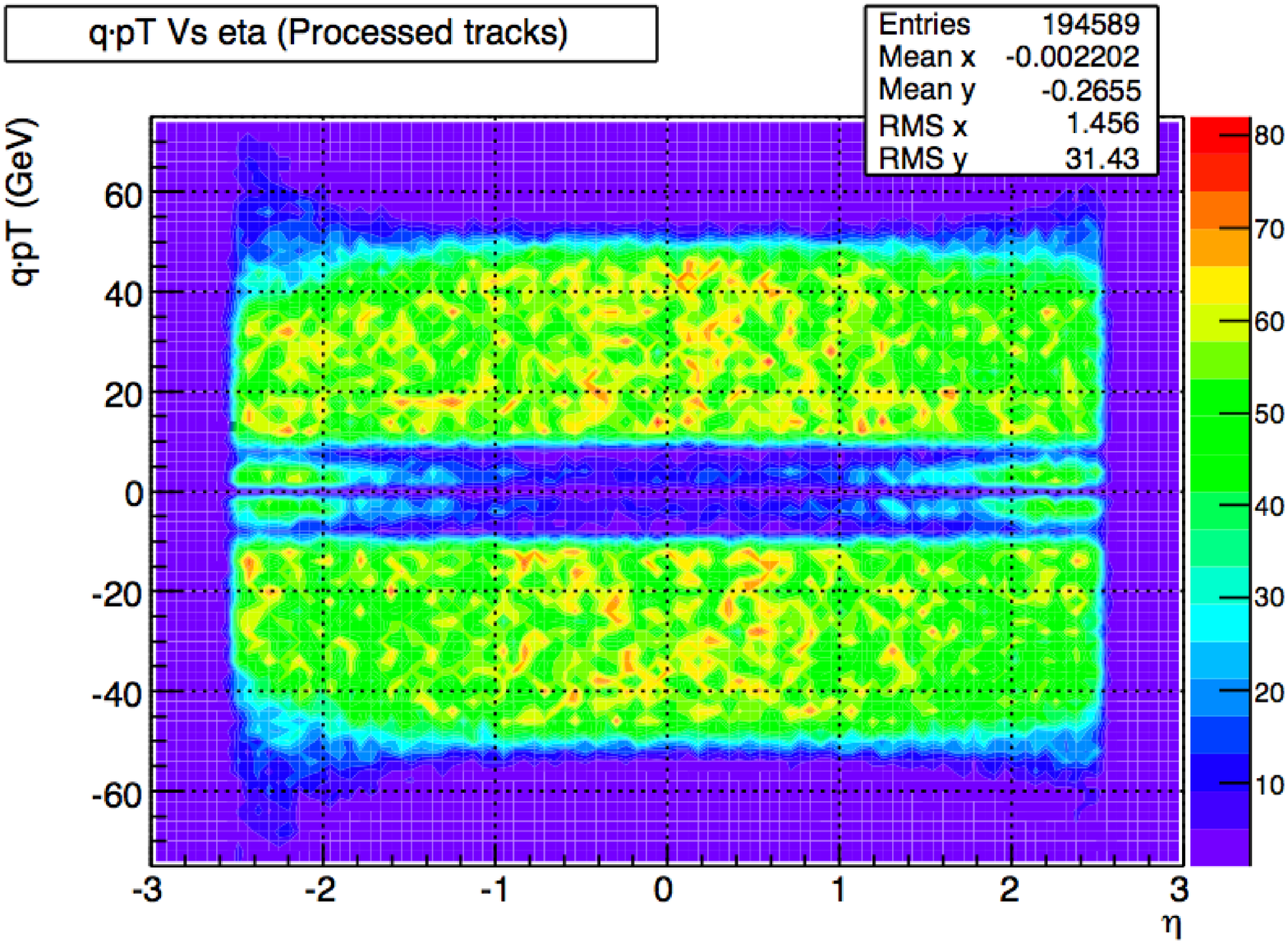}%
\vspace{-2mm}
\caption{\label{qpt_eta}Track q.p$_{T}$ distribution as a function of detector $\eta$ showing improvement from highly misaligned CSC dataset to final FDR results after alignment.}
\end{figure}

Having demonstrated the baseline success of the ID alignment, efforts were concentrated on meeting the demands for the operational model of ATLAS. 
The current model of ID alignment is to be able to provide a first pass alignment constants set and the beam spot position on a daily basis, using dedicated streams. 
A series of week-long mock exercises, the Full Dress Rehearsal (FDR), 
was performed to stress-test the readiness of the alignment platform for the upcoming data.  
The first exercise, FDR1, took place in February 2008. 
By the time of FDR2 (June 2008), the current alignment production model was in place. 
Fig.~\ref{flow} shows the alignment production loop. 
The beam spot calculation and ID alignment were performed in almost automatic mode, using special scripts to steer jobs running in parallel on about 100 CPU's and delivering constants in a 24-hour loop on dedicated computer queues at CERN. The alignment was performed progressively using the Global $\chi^2$  and TRT algorithms, and constants were updated in the database each day upon validation, in increasing levels as mentioned above. 
Simulated event triggers and filters were used to produce an alignment stream (``ID calibration stream") misaligned the same way as the CSC samples. This was extended using a sample of cosmic events with $B$-field. 
At the end of the last exercise in August 2008, the CSC performance was attained in production mode (Fig.~\ref{cosmic_results} and \ref{qpt_eta}). 

\begin{figure}
\includegraphics[width=98mm]{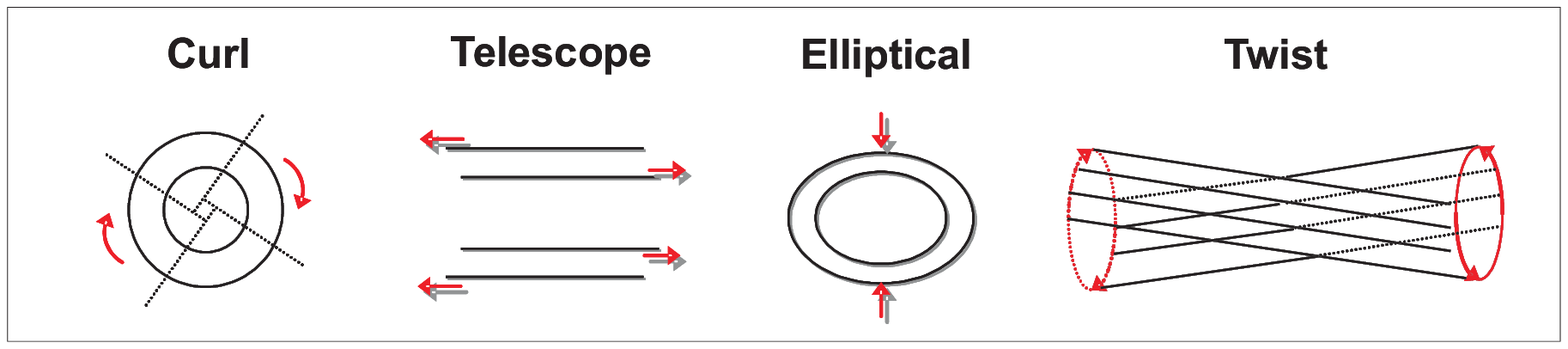}%
\hspace{9mm} \includegraphics[width=40mm]{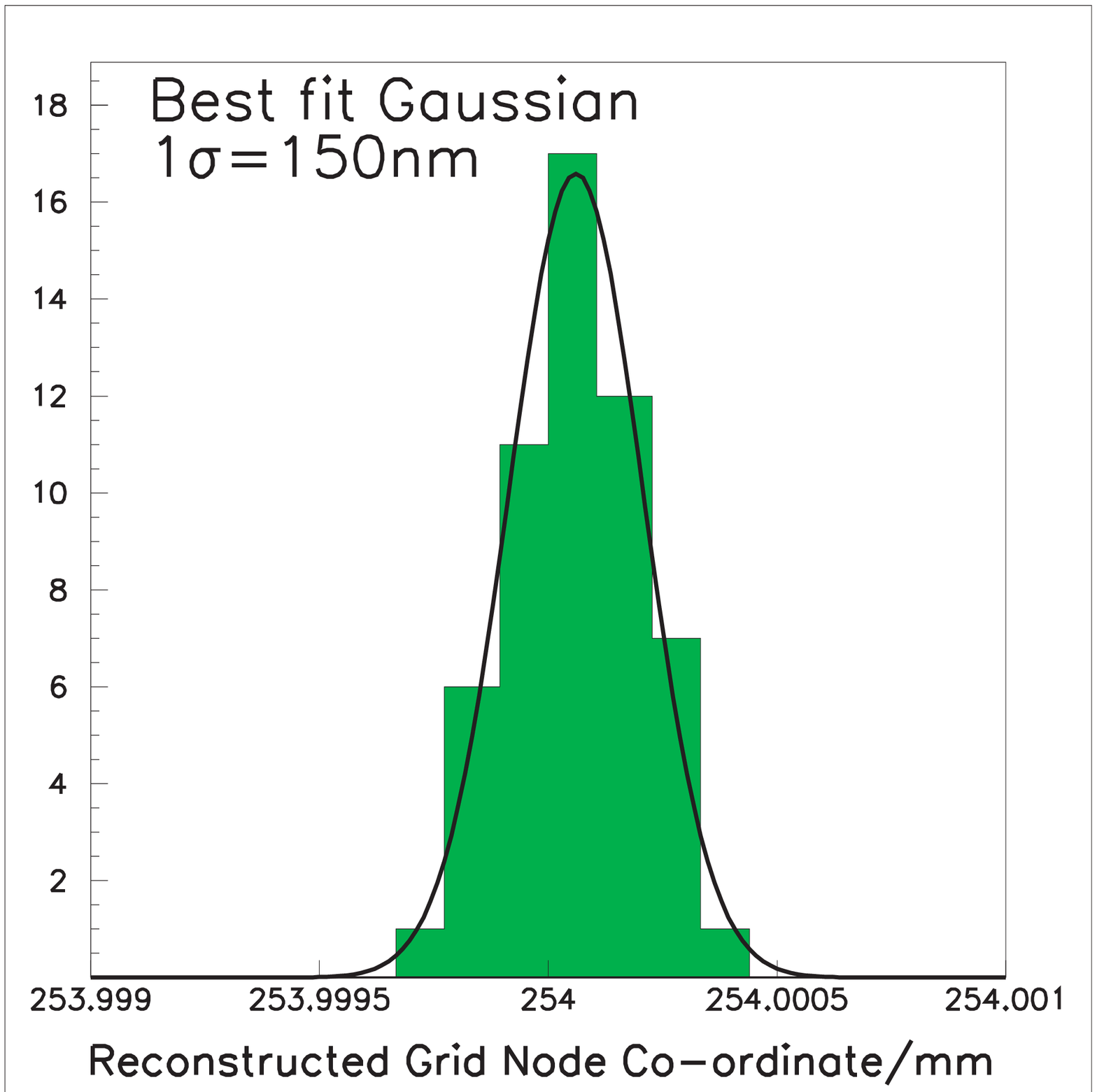}%
\vspace{-1mm}
\caption{\label{3by3}Left: Possible selected distortions of a cylindrical structure which has systematic effects on physics measurements~\cite{not_that_dan_brown}. Right: Demonstrated precision of FSI nodes using small prototype grids.}
\end{figure}

\indent {\em \bf Handling Global Distortions:}
ATLAS will use various constraints to tackle weak mode deformations which bias track measurements. Collision data can be used to constrain some of these modes by imposing a common vertex and using additional information from other subdetectors, such as the muon spectrometer or the calorimeters. 
Cosmic data have already been shown to remove biases in track measurements from radially concentric detector parts (barrels), especially by constraining the $p_T$-biasing modes ~\cite{det_paper}. 
In addition to cosmic rays, a sufficient rate of 
beam halo tracks can help constrain the endcaps by making use of their being parallel to the 
beamline. The study of simulated beam halo and beam gas events for alignment optimisation is ongoing.
Survey data has been incorporated into the alignment code to help initial alignment.
Systematic studies of global distortions with dedicated simulated samples are ongoing for 
the modes believed to have the largest impact on physics (Fig.~\ref{3by3}).

Another external constraint will come  from the Frequency Scanning Interferometry (FSI)~\cite{FSI},
a real-time hardware alignment system for the SCT with all on-detector components installed and connected. 
FSI monitors detector motions every few minutes, using geodetic grids of 842 length measurements 
between nodes on the SCT structure. FSI can reconstruct a 3D grid geometry 
with a precision better than 5 $\mu$m in the critical direction. It has been shown that a 
node precision of 150 nm is achievable (Fig. \ref{3by3}). FSI offers frequent readings of low spatial frequency detector deformations by continuous monitoring, which complements the track-based ID alignment.  
All system interferometer hardware is functioning. Integration of FSI information into the ATLAS offline software is underway.

\section{CONCLUSIONS AND OUTLOOK}

Atlas ID is getting ready for the energetic LHC collisions with an initial alignment being achieved from cosmic ray data using optimised algorithms.
The infrastructure for ID alignment in ATLAS operations mode is in place.
Important systematic effects are being studied using simulations of misaligned detectors.
The online alignment system is fully implemented and is being prepared for continuous beam operation. 


\end{document}